\documentclass[aps,prb,showpacs,twocolumn,citeautoscript,superscriptaddress]{revtex4}
\usepackage{graphicx}
\usepackage{bm}

\newcommand{\nabl}{\bm{\nabla}}

\begin{document}

\title{
Theory of electric polarization in multi-orbital Mott  insulators}

\author{Maxim Mostovoy}
\affiliation{
Zernike Institute for Advanced Materials University of Groningen Nijenborgh 4,
Groningen 9747 AG, The Netherlands
}
\author{Kentaro Nomura}
\affiliation{
Correlated Electron Research Group (CERG), RIKEN-ASI, Wako 351-0198, Japan
            }
\author{Naoto Nagaosa}
\affiliation{
Correlated Electron Research Group (CERG), RIKEN-ASI, Wako 351-0198, Japan
            }
\affiliation{
Cross-Correlated Material Research Group (CMRG), RIKEN-ASI, Wako 351-0198, Japan
}
\affiliation{Department of Applied Physics, The University of Tokyo,
Hongo, Bunkyo-ku, Tokyo 113-8656, Japan}

\date{\today}

\begin{abstract}
The interaction between the electric field, $\mathbf{E}$, and spins in multi-orbital Mott insulators is studied theoretically. We find a generic dynamical coupling mechanism, which works for all crystal lattices and which does not involve relativistic effects. The general form of the coupling is $-T^{ab} E_a e_b$, where ${\mathbf e}$ is the `internal' electric field originating from the dynamical Berry phase of electrons and $T^{ab}$ is a tensor determined by lattice symmetry. We discuss several effects of this interaction:
(i) an unusual electron spin resonance induced by an oscillating electric field,
(ii) the displacement of spin textures in an applied electric field, and
(iii) the resonant absorption of circularly polarized light by Skyrmions, magnetic bubbles, and magnetic vortices.
\end{abstract}

\maketitle


\noindent
{\it Introduction}:
The manipulation of magnetic patterns with an applied electric field is an important issue both for fundamental physics and for applications to spintronics \cite{Bibes_NatMat_2008}. Naively, electrons in Mott insulators should be inert to the electric field oscillations with energies below the charge gap. Yet, a number of multiferroic materials, in which electric polarization is induced by spin orders \cite{Kimura}, show strong response to the electric field at frequencies of magnetic excitations, resulting in the so-called electromagnon peaks in optical absorption \cite{Pimenov_NatMat_2006,Sushkov_PRL_2009}.

Two main mechanisms for the coupling between the electric field and spins have been identified so far~\cite{Katsura_PRL_2005,Mostovoy_PRL_2006,
Sergienko_PRB_2006,Sergienko_PRL2006}.
One is the lattice and electronic polarization induced by the Heisenberg spin exchange energy, which gives rise to the `bond' electric dipoles, ${\mathbf P}_{ij}$, proportional to the scalar products of spins:
 $P^a_{ij} = {\pi}^a_{ij} {\mathbf S}_i \cdot {\mathbf S}_j$ \cite{Sergienko_PRL2006}.
The other originates from the relativistic spin-orbit interaction inducing the dipole moments proportional to the vector products of spins,
${\mathbf P}_{ij} = \alpha {\mathbf e}_{ij} \times ({\mathbf S}_i \times {\mathbf S}_j)$, where ${\mathbf e}_{ij}$ is the unit vector parallel to the bond  \cite{Katsura_PRL_2005,Mostovoy_PRL_2006,Sergienko_PRB_2006}.
The effectiveness of these mechanisms is restricted by symmetry requirements to special lattice geometries and magnetic orders, such as cyloidal spirals and the antiferromagnetic $E$-type order in orthorombically distorted manganites. One of the motivations for this study is to find a generic mechanism that couples electric field to  spin patterns in insulating ferromagnets independently of their crystal structure.

Another motivation is the recent upsurge of interest in Mott insulators close to the transition into metallic state, e.g. $3d$ and $5d$ transition metal oxides \cite{VO2,Kim_Science_2009}, organic crystals \cite{org}, cold atoms \cite{cold}, and quantum dot arrays \cite{dot}. The proximity to metallic state enhances fluctuations of the electron charge density playing the crucial role in the coupling of the low-energy spin degrees of freedom to the electric field. Furthermore, as we show below,  large distances between the localized charges in artificial Mott insulators\cite{dot}, can strongly amplify their response to an applied electric field.

In this paper, we study theoretically the electric polarization induced by time-dependent spin patterns in magnetic insulators taking into account the spin dynamics during the exchange process. We derive the magnetoelectric coupling for a multi-orbital Hubbard model and show that the most universal mechanism that does not require special crystal lattices and relativistic effects gives the electric polarization proportional to the ``internal" electric field ${\mathbf e}({\mathbf x}, t)$
associated with the Berry phase of dynamical spins,
\begin{equation}
{\mathbf e}({\mathbf x}, t) = (1/2) \sin \theta (\partial_t \theta \nabl \varphi - \partial_t \varphi \nabl \theta).
\end{equation}
This quantity has been discussed in the context of the
``electromotive force" or ``spin motive force" in metallic
ferromagnetic systems \cite{Berger,Maekawa,Niu}.
We show that this field is also relevant to the insulating magnets contributing to their dielectric response.

\vskip 0.4cm
\noindent{\it The model}:
Our microscopic model includes the sum of electron Hamiltonians on transition metal sites,
\begin{equation}\label{eq:onsite}
H_{\rm site} =  U n (n-1)+\Delta n_{\beta}
-J_H
\left[\mathbf{S} \cdot (\mathbf{s}_{\alpha}+\mathbf{s}_{\beta})+
\mathbf{s}_{\alpha}\cdot\mathbf{s}_{\beta}\right].
\end{equation}
where the first term is the on-site Coulomb repulsion, $n = n_{\alpha} + n_{\beta}$ being the total number of itinerant electrons on the site, $\Delta > 0$ is energy splitting between the two orbitals $\alpha$ and $\beta$, and the last term is Hund's rule coupling between the local spin $\mathbf{S}_{i}$ and the spin   $\mathbf{s}_{ia} = \frac{1}{2}c_{ia}^{\dagger} \boldsymbol{\sigma} c_{ia}$ of the itinerant electron on the orbital $a=\alpha,\beta$. The zeroth-order Hamiltonian is the sum of all on-site energies and  $H_S$ describing spin interactions that do not originate from exchange processes.

\begin{figure}[b]
\begin{center}
\includegraphics[width=0.46\textwidth]{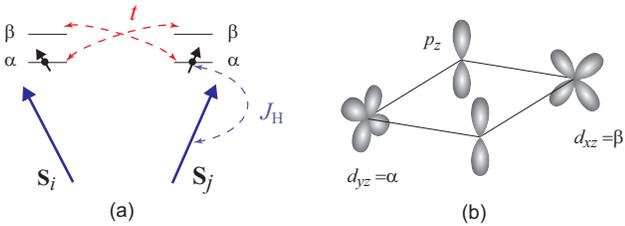}
\caption{
(a) Two-orbital model of electrons interacting with the local spins ${\bf S}_{i}$ and ${\bf S}_{j}$ through the Hund's rule coupling $J_H$.
(b) Physical realization of the two-orbital model describing the hopping between the $\alpha=d_{yz}$ and $\beta=d_{xz}$ orbitals of the magnetic ions on the metal-oxygen plaquette mediated by the $p_z$-orbitals of oxygen ions.}
\label{fig:model}
\end{center}
\end{figure}

We do perturbation theory in the hopping energy of itinerant electrons,
\begin{equation}
V = -\sum_{ia;jb} t_{jb,ia} c^{\dagger}_{jb}c_{ia},
\end{equation}
where $t_{jb,ia} = t_{ia,jb}^{\ast}$ is the amplitude of hopping from the orbital $a$ on the site $i$ to the orbital $b$ on the site $j$. In the presence of
electric field the hopping amplitudes are modified using the `Peierls substitution'~\cite{Peierls_ZPhys_1933},
\begin{equation}\label{eq:Peierls}
t_{jb,ia} \rightarrow
t_{jb,ia}e^{-i e \int_{\mathbf{x}_i}^{\mathbf{x}_j}d\mathbf{x}\cdot
\mathbf{A}(\mathbf{x},t)},
\end{equation}
where $\mathbf{A}$ is the vector potential, $-e$ is the electron charge and $\hbar = c = 1$ (furthermore, the term
$-e A_{0}n$ is added to the on-site Hamiltonian).

In general, the electric field dependence of hopping amplitudes is more complex than the one described by Eq.(\ref{eq:Peierls}). If the metal-ligand-metal bond has a nonzero electric dipole moment,  an applied electric field parallel to the dipole results in a linear dependence of the hopping amplitude on $\mathbf{E}$, which gives rise to the exchange striction and relativistic mechanisms discussed in the introduction \cite{Katsura_PRL_2005,Sergienko_PRB_2006,Mostovoy_PRL_2006,Sergienko_PRL2006}.
Here we only take into account the electric field dependence resulting from the Peierls substitution present for any bond geometry.

The model with two orbitals per site is the simplest model of a multi-orbital Mott insulator. We first show that the electron hopping between different orbitals, e.g. the orbital $\alpha$ on site 1 and the orbital $\beta$ on site 2 (see Fig.~\ref{fig:model}), favoring ferromagnetic interactions between spins, gives rise to a dynamical electric polarization induced by rotating spins. Then we  explain why this does not happen in the single-orbital Hubbard model. For simplicity we assume that there is only one itinerant electron per site and consider the strong Hund's rule coupling and large $S$ limit, in which the spin of the itinerant electron is parallel to the local spin.

\noindent
{\it Electric polarization of spins}:
First we discuss interactions between two sites, $1$ and $2$, in the ferromagnetic model, in which only
$t_{j\alpha,i\beta} = t_{i\alpha,j\beta} = t \neq 0$ (see Fig.~\ref{fig:model}). To second order in the hopping amplitude $t$ the correction to the imaginary
time spin action is
\begin{equation}
\delta S = t^2 \int_{0}^{\beta} d\tau_{\rm i}
\int_{\tau_{\rm i}}^{\beta} d\tau_{\rm f} e^{-U'(\tau_{\rm f}-\tau_{\rm i})}
\left[C_{21} c_{21} + C_{12} c_{12}\right],
\label{eq:correction}
\end{equation}
where $U' = U + \Delta - J_{H}/4$,
\begin{eqnarray}\label{eq:c21}
C_{21} &=& e^{-ie\int_{\tau_{\rm i}}^{\tau_{\rm f}}d\tau
\int_{\mathbf{x}_1}^{\mathbf{x}_2}d\mathbf{x} \cdot \mathbf{E}}, \nonumber \\ \\
c_{21} &=& e^{\int_{\tau_{\rm i}}^{\tau_{\rm f}}d\tau
\left[
\langle \mathbf{n}_1(\tau)|\partial_{\tau}|\mathbf{n}_1(\tau)
\rangle
-
\langle \mathbf{n}_2(\tau)|\partial_{\tau}|\mathbf{n}_2(\tau)
\rangle
- \delta H_{21}
\right]}
\nonumber \\
&~&
\times\langle \mathbf{n}_1(\tau_{\rm f})\vert\mathbf{n}_2(\tau_{\rm f})
\rangle
\langle \mathbf{n}_2(\tau_{\rm i})\vert\mathbf{n}_1(\tau_{\rm i})
\rangle, \nonumber
\end{eqnarray}
with $\delta H_{21} = \frac{1}{2(S+1/2)}\left(
\mathbf{n}_2 \!\cdot\! \frac{\partial}{\partial \mathbf{n}_2}
- \mathbf{n}_1 \!\cdot\! \frac{\partial}{\partial \mathbf{n}_1}\right)H_{S}$ describing the spin energy change due the electron hopping from site 1 to site 2 ($C_{12},c_{12}$ are obtained by interchanging the indices $1$ and $2$). Here,  $\mathbf{n}_{1,2}$ is the unit vector in the direction of the local spin $\mathbf{S}_{1,2}$ and $\vert\mathbf{n}\rangle$ denotes the eigenstate of the itinerant electron with spin parallel to $\mathbf{n}$.

$C_{21}$ is invariant under local gauge transformations and so is $c_{21}$. Introducing the vector potential of the `internal' field by
$
a_0 =
i \langle
\mathbf{n}\vert\partial_{\tau}|\mathbf{n}
\rangle$ and
$
\mathbf{a} =
i \langle
\mathbf{n}\vert\partial_{\mathbf{x}}|\mathbf{n}
\rangle
,$
we can write the overlap of the electron spin wave functions in the form,
\begin{equation}
\langle \mathbf{n}_2\vert\mathbf{n}_1
\rangle = \cos \frac{\theta_{21}}{2}
e^{ i \int_{\mathbf{x}_1}^{\mathbf{x}_2}d\mathbf{x}\cdot
\mathbf{a}},
\end{equation}
%
%
where $\theta_{21}$ is the angle between $\mathbf{n}_1$ and $\mathbf{n}_2$.
The vector $\mathbf{n}(\mathbf{x})$ that defines $\mathbf{a}$ varies between
$\mathbf{n}_1$ to $\mathbf{n}_2$ along the segment of a circle on the unit sphere,
while $\mathbf{x}$ varies from $\mathbf{x}_1$ to $\mathbf{x}_2$.
We now can write $c_{21}$ in the manifestly covariant form
\begin{equation}
c_{21} = \cos \frac{\theta_{21}(\tau_{\rm f})}{2}
\cos \frac{\theta_{21}(\tau_{\rm i})}{2}
e^{\int_{\tau_{\rm i}}^{\tau_{\rm f}}d\tau
(i\int_{\mathbf{x}_1}^{\mathbf{x}_2}d\mathbf{x}\cdot \mathbf{e}-\delta H_{12})},
\label{eq:c21e}
\end{equation}
where $\mathbf{e} = \partial_{\mathbf{x}} a_0 - \partial_{\tau}\mathbf{a}$
is the gauge invariant internal electric field.
Comparing Eqs.(\ref{eq:c21}) and (\ref{eq:c21e}), we find that the correction to action only depends on the combination of applied and internal electric fields,
$e\mathbf{E} - \mathbf{e}$.

Since the time spent by the hopping electron on a neighboring site,
$\tau_{\rm f} - \tau_{\rm i} \sim (U')^{-1}$, is much shorter than the characteristic time of spin dynamics, $C_{ij}$ and $c_{ij}$ in Eq.(\ref{eq:correction}) can be expanded in powers of $\tau_{\rm f} - \tau_{\rm i}$, which generates an expansion of the spin action in powers of $(U')^{-1}$. To lowest order we obtain an effective ferromagnetic
interaction between the spins,
\begin{equation}
H_{\rm eff} = - \frac{t^2}{U'}\left(\mathbf{n}_1\cdot\mathbf{n}_2 + 1\right).
\end{equation}
The third-order term in the expansion gives the interaction described by the real time Lagrangian,
\begin{equation}
L = \frac{t^2}{(U')^3}\left(\mathbf{n}_1\!\cdot\!\mathbf{n}_2 + 1\right)\!\left[\int_{\mathbf{x}_1}^{\mathbf{x}_2}\!\!d\mathbf{x}\!\cdot\!\left(e\mathbf{E} - \mathbf{e}\right) + \delta H_{21}\right]^2.
\end{equation}
The term $\propto \mathbf{E}^2$ is the spin contribution to the dielectric susceptibility, while the term linear in $\mathbf{E}$ describes the coupling of the external electric field to the spin-induced electric polarization:
\begin{equation}\label{eq:coupling}
L_{E} = {\cal E} \left\{\left({\dot \mathbf{n}}_1 + {\dot \mathbf{n}}_2\right)\cdot \left[\mathbf{n}_{1} \times \mathbf{n}_{2}\right] + 2\left(\mathbf{n}_1\cdot\mathbf{n}_2 + 1\right)\delta H_{21}\right\},
\end{equation}
where ${\cal E} = \frac{t^2e\mathbf{E} \cdot (\mathbf{x}_2-\mathbf{x}_1)}{(U')^3}$. The second term in curly brackets describes the charge re-distribution between the two sites that takes place when $\mathbf{n}_1 \neq \mathbf{n}_2$, while the first term is the dynamical polarization originating from the coupling between the external and internal electric fields, $\mathbf{E}\cdot\mathbf{e}$, which for insulators was not considered before. Though weak, it leads to a number of unusual effects discussed below.

\noindent
{\it Physical consequences}:
The dynamical part of $L_{E}$ is eliminated by the rotation of the spins around $\left[\mathbf{n}_{1} \times \mathbf{n}_{2}\right]$,
\begin{equation}
\left\{
\begin{array}{ccc}
\delta \mathbf{n}_1 &=& \frac{{\cal E}}{(S+1/2)}
\left(\mathbf{n}_2 - (\mathbf{n}_1\cdot\mathbf{n}_2) \mathbf{n}_1\right),\\ \\
\delta \mathbf{n}_2 &=& -\frac{{\cal E}}{(S+1/2)}
\left(\mathbf{n}_1 - (\mathbf{n}_1\cdot\mathbf{n}_2) \mathbf{n}_2\right),
\end{array}
\right.
\end{equation}
applied to the Berry-phase term in the spin Lagrangian,
\begin{equation}
L_B = (S + \frac{1}{2}) \sum_{i = 1,2} \left(\cos \theta_i - 1\right) {\dot \varphi}_i,
\end{equation}
where $\theta_i$ and $\varphi_i$ are the polar angles describing $\mathbf{n}_i$. If the spin Hamiltonian, $H_S$,
is not rotationally invariant, this transformation generates a coupling to electric field in the spin Hamiltonian. For example, the magnetic anisotropy,
\begin{equation}
H_S = \frac{A}{2} \left[ \left({\tilde S}_1^z\right)^2+\left({\tilde S}_2^z\right)^2\right],
\end{equation}
where  ${\tilde \mathbf{S}}_i = \mathbf{S}_i + \mathbf{s}_{i\alpha}$, gives rise to the interaction
\begin{equation}\label{eq:Hint}
H_{\rm int} = \frac{A {\cal E}(t)}{{\tilde S}}
\left[ \left({\tilde S}_1^z\right)^2-\left({\tilde S}_2^z\right)^2\right]
\end{equation}
(${\tilde S} = S+ \frac{1}{2}$).  The time-dependent electric field induces transitions between the states of opposite parity under the permutation of the two sites, corresponding to even and odd values of the total spin of the two sites, $S_t$, while the projection of the total spin on the anisotropy axis, $S_t^z$, remains constant. These selection rules have to be compared with those for the conventional ESR where the time-dependent magnetic field leaves the total spin unchanged, while its projection on the static magnetic field changes by $\pm 1$.

Consider now an arbitrary spin texture in a ferromagnet with $\mathbf{n}(\mathbf{x},t)$ varying slowly at the lattice constant scale. The Lagrangian describing the linear coupling of the texture to the electric field is given by
\begin{eqnarray}
L_{E} &=& - \int\!\!d^3x T^{ab} E_a e_b(\mathbf{x},t), \;\;\;\;\;\mbox{where}\nonumber\\\label{eq:continuum} \\
T^{ab} &=& \frac{e}{(U')^3} \frac{1}{v} \sum_j
|t_{j\beta,i\alpha}|^2 (x_{j}^a - x_{i}^a)(x_{j}^b - x_{i}^b),\nonumber
\end{eqnarray}
($v$ is the unit cell volume). For a simple cubic lattice with the lattice constant $a$, $T^{ab} = \frac{g}{a^3} \delta^{ab}$ with $g = \frac{2ea^2t^2}{(U')^3}$. Importantly, Eq.(\ref{eq:continuum}) entirely comes from the dynamical part of Eq.(\ref{eq:coupling}), as the static polarization in the continuum limit is a total derivative.

Similarly to the two-spin case, the transformation  $\mathbf{n}(\mathbf{x},t) \rightarrow \mathbf{n}(\mathbf{x}+\mathbf{X},t)$ with $\mathbf{X} = \frac{g}{{\tilde S}}\mathbf{E}$, applied to the Berry phase term, cancels the interaction Eq.(\ref{eq:continuum}). Since this transformation leaves the Hamiltonian of a translationally invariant system unchanged, the effect of electric field is to shift a spin texture as a whole by the vector $- \mathbf{X}$. The shift is a small fraction of the lattice constant: $\frac{X}{a} \sim \frac{t^2 eEa}{(U')^3}$. It can, however, be enhanced by proximity to the metal-insulator transition (through larger $t/U$ ratio) and by a larger distance between the spins, $a$, in magnetic semiconductors and quantum dot arrays.

A much stronger effect of the interaction Eq.(\ref{eq:continuum}) is the resonant absorption of circularly polarized light by Skyrmions, magnetic bubbles, and magnetic vortices. These magnetic defects in two spatial dimensions carry a nonzero topological charge \cite{Polyakov_JETPLett_1975}, $Q = \frac{1}{4\pi} \int \!\! d^2x \mathbf{n} \cdot \partial_x \mathbf{n} \times \partial_y \mathbf{n}$, integer for Skyrmions/bubbles and half-integer for vortices.

Magnetic vortices are spontaneously induced by magnetostatic interactions in nanodiscs of ferromagnetic metals
\cite{Wachowiak_Science_2002}, while periodic arrays of magnetic bubbles appear in thin-film
ferromagnets with a strong out-of-plane anisotropy upon application of magnetic field on the order of $10^2$ Oe \cite{Malozemoff_1979}. Similar arrays of skyrmions, which are bubbles with ``thick" domain walls, were recently observed in bulk ferromagnetic metals without inversion symmetry \cite{Muehlbauer_Science_2009,Yu_Nature_2010}.

According to Eq.(\ref{eq:continuum}), a moving topological defect induces the net electric dipole moment in the direction normal to its velocity, $\mathbf{P} \propto  g Q  \left[{\hat \mathbf{z}} \times {\dot \mathbf{R}} \right]$, where
$\mathbf{R} = (R_x,R_y)$ is the position of the center of the defect and ${\hat \mathbf{z}}$ is the unit vector normal to the film. Consider such a defect in a ferromagnetic insulator with a confined geometry, which breaks translational symmetry, e.g. a vortex in a nanodisc. We assume that the confining potential has the form, $U = \frac{K}{2} (R_x^2+R_y^2)$. In the adiabatic limit the dynamics of the collective coordinates $R_i$ is described by Thiele equations\cite{Thiele_PRL_1973,Huber_PRB_1982,Tretiakov_PRL_2008}
\begin{equation}
G_{ij} \left({\dot R}_j + \frac{g}{{\tilde S}} {\dot E}_j\right) + \alpha \Gamma_{ij} {\dot R}_{j} = - \frac{\partial U}{\partial R_i},
\label{eq:Thiele}
\end{equation}
to which we added the coupling of spins to the electric field. Here, $\alpha$ is the Gilbert damping constant and the nonzero components of the tensors $G_{ij}$ and $\Gamma_{ij}$ are,
\begin{equation}
\begin{array}{ccccl}
G_{xy} &=& - G_{yx} &=& 4 \pi Q \;\;\;\mbox{and}\\ \\
\Gamma_{xx} &=& \Gamma_{yy} &=& \int \!\! d^2x \partial_i \mathbf{n} \cdot \partial_i \mathbf{n}.
\end{array}
\end{equation}
In absence of electric field the (damped) eigenmode
$\mathbf{R}(t) \propto \left(\cos\Omega t, -q\sin\Omega t\right)$ describes the
rotational motion of the center of the spin texture with the frequency $\Omega = \frac{K}{4 \pi |Q|}$
in the direction defined by $q = \mbox{sign}(Q) = \pm 1$. The response to the rotating electric field
\begin{equation}
\mathbf{E}(t) = E_{\omega} \left(\cos \omega t,-\sigma \sin\omega t\right),
\end{equation}
($\sigma = \pm 1$) at the resonant frequency, $\omega = \Omega$, is given by
\begin{equation}
X_{\Omega} =
\frac{g E_{\Omega}}{2{\tilde S}}
\left\{
\begin{array}{rl}
\frac{i}{\Omega \tau}, & \mbox{for $\sigma = + q$},\\ \\
-\frac{1}{2 - i\Omega \tau}, & \mbox{for $\sigma = - q$,}
\end{array}
\right.
\end{equation}
where $\tau = \frac{\alpha \Gamma_{xx}}{K}$ is the relaxation time. For $\Omega \tau \sim \frac{\alpha}{4\pi} \ll 1$ the excitation of the rotational motion by the electric field with $\sigma = + q$ is resonantly enhanced by the factor $\frac{1}{\Omega \tau}$ compared to the shift in translationally invariant systems, while for $\sigma = - q$ there is no enhancement. For magnetic insulators with $\alpha \sim 10^{-3}-10^{-2}$ (see e.g. Ref.~\onlinecite{Malozemoff_1979}) the resonant enhancement by three to four orders of magnitude opens a possibility to manipulate spin textures with an electric field.

\noindent
{\it Discussion: }Now we discuss why the presence of several orbitals is essential to obtain the linear coupling of spins to electric field  Eq.(\ref{eq:coupling}) and why such a coupling does not exist in the single-orbital model, in which only $t_{j\alpha,i\alpha} \neq 0$ and interactions between spins are antiferromagnetic. In that case
\begin{eqnarray}
c_{21} &=& e^{\int_{\tau_{\rm i}}^{\tau_{\rm f}}d\tau
\left[
\langle \mathbf{n}_1(\tau)|\partial_{\tau}|\mathbf{n}_1(\tau)
\rangle
-
\langle -\mathbf{n}_2(\tau)|\partial_{\tau}|-\mathbf{n}_2(\tau)
\rangle
\right]}
\nonumber \\
&~&
\times \langle \mathbf{n}_1(\tau_{\rm f})\vert-\mathbf{n}_2(\tau_{\rm f})
\rangle
\langle -\mathbf{n}_2(\tau_{\rm i})\vert\mathbf{n}_1(\tau_{\rm i})
\rangle
\end{eqnarray}
(we ignore the changes in the magnetic energy $\delta H_{21}$, since the static polarization disappears in the continuum limit). We note that for a particular choice of
spinor wave functions, describing electrons with spin parallel/antiparallel to the local spin,
\begin{equation}
\vert\mathbf{n}
\rangle
=
\left(
\begin{array}{c}
\cos \frac{\theta}{2} \\
\sin \frac{\theta}{2} e^{i \varphi}
\end{array}
\right),\;\;\;\;
\vert - \mathbf{n}\rangle
=
\left(
\begin{array}{c}
-\sin \frac{\theta}{2} e^{-i \varphi}  \\
\cos  \frac{\theta}{2}
\end{array}
\right),
\end{equation}
$c_{21}$ is symmetric with respect to the interchange of  the indices $1$ and
$2$: $c_{12} = c_{21}$. Since the $\mathbf{E}$-dependent term, $C_{21}$, is antisymmetric
with respect to this permutation, there is no linear coupling of spins to electric
field. Due to the gauge invariance of $c_{21}$ this result is independent of
choice of spinor wave functions.

This result can be physically explained as follows. The first term in  Eq.(\ref{eq:coupling})originates from the spin dynamics in the virtual states with two electrons occupying the same site. More precisely, it describes the difference between the spin rotations in the state where both electrons occupy site 1 and the state with two electrons on site 2. In the single-orbital case, however, the virtual state of two itinerant electrons is a spin-singlet independent of which site is doubly occupied and which is empty. Hence, no dynamical linear coupling to electric field.

In multi-orbital Mott insulators both exchange processes described by the models I and II take place. Importantly, the coupling of spins to the electric field, expressed by Eqs.~(\ref{eq:coupling}) and (\ref{eq:continuum}), is present  even if the dominant interaction between spins is antiferromagnetic, i.e. our results apply to insulators with ferrimagnetic, canted and spiral spin orders.

There are several interesting problems left for future studies. One is the relevance of the present mechanism for spin liquid states~\cite{NgLee}, where effects of the spin Berry phase are enhanced by strong spin fluctuations.  Another issue of interest is the ring-exchange processes giving rise to persistent orbital currents in Mott insulators~\cite{orb}. Dynamical effects resulting from the ring-exchange process deserve a scrutiny.

In summary, we show that the dynamical spin Berry phase in multi-orbital Mott insulators couples the electric field to the translational modes of spin textures. We derive equations of motion for the center-of-mass coordinates of Skyrmions and magnetic vortices in an applied electric field and predict  the resonant absorption of the circularly polarized light by these topological objects as well as the ESR effect where spin transitions are induced by the time-dependent electric field.

We gratefully acknowledge fruitful discussions with S. Maekawa, Q. Niu, Y. Tokura,
Yu  Xiuzhen, and Y. Onose. This research is supported by MEXT Grand-in-Aid
No.20740167, 19048008, 19048015 and, and 21244053, Strategic International Cooperative
Program (Joint Research Type) from Japan Science and Technology Agency, and by
the Japan Society for the Promotion of Science (JSPS) through its ``Funding Program for
World-Leading Innovative R \& D on Science and Technology (FIRST Program)''. MM is grateful for hospitality at RIKEN.

\end{document}